# Fault Diagnosis of Inter-turn Short Circuit in Permanent Magnet Synchronous Motors with Current Signal Imaging and Unsupervised Learning

Wonho Jung
*Department of Mechanical Engineering*
*Korea Advanced Institute of Science and Technology*
Daejeon, South Korea
wonho1456@kaist.ac.kr

Sung-Hyun Yun
*Department of Mechanical Engineering*
*Korea Advanced Institute of Science and Technology*
Daejeon, South Korea
tifon97@kaist.ac.kr

Yoon-Seop Lim
*Department of Mechanical Engineering*
*Korea Advanced Institute of Science and Technology*
Daejeon, South Korea
bu2050@kaist.ac.kr

Sungjin Cheong
*Department of Mechanical Engineering*
*Korea Advanced Institute of Science and Technology*
Daejeon, South Korea
sjcheong21@kaist.ac.kr

Jaewoong Bae
*Department of Mechanical Engineering*
*Korea Advanced Institute of Science and Technology*
Daejeon, South Korea
jwoong.bae@kaist.ac.kr

Yong-Hwa Park
*Department of Mechanical Engineering*
*Korea Advanced Institute of Science and Technology*
Daejeon, South Korea
yhpark@kaist.ac.kr

*Abstract*—This paper proposes machine-independent feature engineering for winding inter-turn short circuit fault that uses electrical current signals. Electrical current signal collected from permanent magnet synchronous motor (PMSM) is subjected to different environmental and operational conditions. To solve these problems, robust current signal imaging method and deep learning-based feature extraction method are developed. The overall procedure includes the following three key steps: (1) transformation of a time-series current signal to two-dimensional image, (2) extracting features using convolutional neural networks, and (3) calculating a health indicator using Mahalanobis distance. Transformation of the time-series signal is based on recurrence plots (RP). The proposed RP method develops from feature engineering that provides the dominant fault feature representations in a robust way. The proposed RP is designed that maximizes the features of inter-turn short fault and minimizes the effect of noise from systems with various capacities. To demonstrate the validity of the proposed method, two case studies are conducted using an artificial fault seeded testbed with two different capacities of motor. By calculating the feature using only the electrical current signal of the motor without the parameters related to the capacity of the motor, the proposed feature can be applied to motors with different capacities while maintaining the same performance.

## I. Introduction

Permanent magnet synchronous motors (PMSM) are widely used in industry applications such as power plant, manufacturing lines, high-speed trains, and electric vehicles. Unexpected faults of PMSM as a key component in most of the engineered systems can lead to significant financial, societal, and human losses. It is important to diagnose PMSM faults so that the losses can be minimized. Failure from PMSM causes an unbalance in the three-phase current of the electric motor and degrades the efficiency and performance of the electric motor [1,2]. Therefore, electrical characteristics such as resistance and impedance of motor stator can be used for motor fault diagnosis. However, electrical signals respond to very slight signal changes and are difficult to detect. To solve this problem, many fault diagnosis method using a vibration sensor are already conducted, however, vibration sensors are limited to use in the real industrial field due to the difficulties of installing the vibration sensor [3,4].

Fault diagnosis method can be categorized into three types: model-based, data-driven, and signal based method [5, 6]. First, the model-based fault diagnosis method derives the target system based on domain knowledge and estimates parameters relate to the faults [7,8]. If the target system is complex, a lot of time and effort are required in the system modeling process. In addition, real-time fault detection is impossible due to high computational cost. Second, the data-driven method uses information contained in a large amount of data measured from the target system [9]. Compared with the model-based method, domain knowledge is not required much. Therefore, the data-driven method is effective for fault diagnosis of the complex engineered system. Recently, the data-driven method has been widely studied with machine learning and artificial intelligence technologies. For developing a high-performance fault diagnosis method, a lot of healthy data and faulty data to be diagnosed are needed. However, it is difficult to collect a lot of fault data due to the regular maintenance and the continuous improvement of the system. Finally, signal-based method diagnoses the fault by analyzing the signals associated with the fault [10, 11]. The types of signals to be measured are determined by the operating environment and the types of faults. The types of signals used for fault diagnosis are vibration, current, sound, pressure, and temperature. A feature that indicates the health state of the target system can be designed by analyzing signals measured from the system in time, frequency, and time-frequency domains based on domain knowledge [12-14]. Typical analysis methods include fast Fourier transform (FFT), short-time Fourier transform (STFT), and wavelet transform method. However, research that detects the weak change of signal reflecting the failure characteristics is conducted limitedly.

To fill this research gap, this paper proposes the time-series imaging method that can detect weak change of signal. The proposed time-series imaging method can be applied the PMSM only with the motor current signal and without the specific parameters of the motor for fault diagnosis. The overall procedure includes the following three key steps: (1) transformation of a one-dimensional current signal to a two-dimensional image in time-domain, (2) extracting features using convolutional neural networks, and (3) calculating a health indicator using Mahalanobis distance. The proposed method includes a process of modifying the recurrence plot (RP). In the modifying process, this paper pursues maximizing the weak fault features and minimizing the noise effects with various capacities of PMSM. The proposed method is validated by electric current signal measured from motor testbed.

The remainder of this paper is organized as follows. Section II provides an overview of the theoretical background of the motor current signature analysis (MCSA) method and time-series imaging method. Section III provides the proposed method including the proposed time-series imaging method, feature extraction, and health indicator. To validate the effectiveness of the proposed method, in Section IV, a case study is described that uses the artificial fault seeded testbed. Finally, Section V concludes the paper with a discussion of future work.

## II. A Brief Overview of Related Works

### A. Motor Current Signature Analysis

As the mechanical systems already use current signals to control motors, MCSA method has the advantage that no additional sensors are required for fault diagnosis. Therefore, MCSA has been utilized in many motor failure diagnosis studies [5].

There are two approaches exist in MCSA method. First, analyze the frequency domain in motor current signals. Kim et al. [15] proposed the fault diagnosis method using the fault frequency of the electric current signal. The health indicator is defined as the magnitude of third harmonic frequency at faulty state divided by the magnitude of third harmonic frequency at heathy state. Even if the fault has the same severity, the magnitude of the third harmonic frequency can be change due to the operating condition. After recording the magnitude of third harmonic frequency under certain conditions, the needed value is calculated through interpolation. When the health indicator exceeded the threshold, value defined in advance by the experiment, it means that the system has the inter-turn short circuit fault. The advantage of this method is that a principle of fault diagnosis is interpretable because it is based on domain knowledge. On the other hand, the disadvantage of this method using the characteristic frequency is collecting the electric current signal for a long time because it requires a high resolution. Also, a digital signal processing technique optimized for a specific PMSM is required in order to accurately calculate the characteristic frequency. Second, there is also a fault diagnosis method using the electric current signal characteristic in time domain. J. F. Martins et al. [16] diagnosed the fault using the characteristic whose electric current of three-phase is unbalanced when inter-turn short circuit fault occurs. A graphical representation of the transformed alpha-beta current by a Clarke-Concordia transform shows a circle in case of the healthy state and an ellipse in case of the faulty state. The degree of distortion was used as a health indicator for the diagnosis of inter-turn short circuit fault. This health indicator has the advantage that domain knowledge is not required and it is simple because it does not use FFT. However, there is a disadvantage that it is not possible to compare the fault severity because the degree of distortion varies depending on the characteristic of the PMSM.

Several features can be used in the frequency domain and time domain to increase the performance of fault diagnosis method. A. Widodo et al. [17] studied the fault diagnosis method using several features extracted from the electric current and vibration signal measured from the testbed. Seven features such as mean, root means square, skewness, and kurtosis extracted from the vibration signal and five features such as crest factor and entropy error were extracted from the electric current. The high dimensional features were converted into a low dimensional features using kernel principal component analysis. The converted features were classified with high performance for six fault modes using support vector machine. Since the fault data of the target system are required for the learning of the classifier, applying the classifier trained for the specific PMSM to the other PMSM cannot expect high performance. Therefore, for realistic fault diagnosis research, it is necessary to focus on the feature extraction of motor-specific faulty characteristic factors regardless of motor specifications.

### B. Time-Series Imaging Method

Time-series data are valuable data for diagnosis because it is a sequence of data points in natural temporal ordering. However, time-series data analysis is a quite difficult task due to the preliminary investigation of various features and complicated feature extraction. Recently, convolutional neural network (CNN) has achieved good performance for the prognostics and health monitoring (PHM) field. Unlike traditional feature engineering, CNN does not require hand-crafted features. Most CNN mainly uses 2D image data due to the comprehensive information. However, most of the mechanical industries gather 1D time-series data instead of 2D image data, the time-series imaging is one of the challenging researches for deep learning.

In many recent years, a number of methods have been devised to transforming 2D images from time-series data such as Gramian angular field (GAF), Markov transition field (MTF) [18], dynamic time warping (DTW) [19], and recurrence plot (RP) [20]. GAF is an image extracted from a time-series, representing some temporal correlation between each time data point. The background of GAF represents time-series in a polar coordinate system instead of the typical Cartesian coordinates. However, GAF has a slow conversion speed because it should rescale their time data points before converting to image, and lots of calculations are needed. MTF represents the Markov transition probabilities sequentially to preserve information in the time domain. Markov transition matrix is insensitive to the distribution of time-series data and temporal dependency on time steps, but the Markov transition matrix should be predefined. DTW is one of the algorithms for measuring similarity between two temporal sequences, which may vary in speed. These similarity measures can be interpreted as the probability that two points are related. DTW is not affected by the length of the signal, but other signals are necessary for measuring similarity. On the contrary, RP is calculated on a spatial trajectory of time-series data without any operator. RP describes a collection of time pairs whose orbits are at the same location. They reveal different patterns of RP for time-series with randomness, periodicity, chaos, and trend. RP works relatively fast compared to other imaging methods and reflects the various characteristics of time-series data.

## III. Proposed Method

This section proposes a current signal imaging method based on modified RP, and a new health indicator for diagnosing the inter-turn short circuit fault. The proposed approach includes three key steps: (1) time-series data imaging method using modified RP in time-domain, (2) extracting features using CNN, and (3) calculating a health indicator using Mahalanobis distance (MD). Section III-A describes current signal imaging method. Sections III-B introduce the health indicator using CNN and MD.

### A. Current Signal Imaging Method using Modified RP

RP provides an easy way to visualize the periodic information of a trajectory through a phase domain. RP includes two steps: (1) find the m-dimensional spatial trajectory of time series data, (2) find the distance matrix from the m-dimensional trajectories of the time series. Therefore, RP can contain all relevant dynamical information of time-series data and display diagonally with all-time pairs separated by multiples of time-series data cycle. However, original RP has difficulty detecting weak data changes because the value of RP is fixed to a one or zero according to the predefined threshold. Most fault signals in time domain are very weak and easily affected by external noise. Therefore, proposed RP can be formulated as

$$R_p(i,j) = \begin{cases} N & \|x(i)-x(j)\| \geq N \\ \|x(i)-x(j)\| & \text{otherwise} \end{cases} \quad (1)$$

where $i$ and $j$ are the vectors of the time-series data points, respectively; $N$ is a predefined threshold ($N>0$) and $R_p$ is transformed proposed RP matrix. Proposed RP equation represents that the big data change in time domain is displayed by specific threshold value $N$ and the weak data change in time domain is displayed by the weak data change value directly. Besides, proposed RP works robustly in noisy conditions by mapping their data value if under specific threshold value $N$. To emphasize weak signals, proposed RP is reconstructed by sparse dictionary learning. Dictionary learning is a branch of signal processing and machine learning that aims at finding a dictionary in which some training data admits a sparse representation. Finding sparse components means finding key points that classify labels among training data. Therefore, the sparser the representation, the better the dictionary. In this paper, the image pattern is further emphasized by multiplying the proposed RP image using the dictionary component constructed through sparse learning. Several time domain features for condition monitoring are already reviewed [17]. However, there are still unknown features in time-series data. This time-series data easily can be contaminated by noise. Therefore, proposed RP is a useful tool for process unknown dynamic characteristics features.

### B. Health Indicator with Mahalanobis Distance and CNN

CNN is artificial neural networks that incorporate convolution operations. By using the convolution operations, the spatial information of the 2D data can be maintained and sent to the next layer. The initial layers of CNN learn basic low-level image features such as edges, while the deeper layers learn more complicated features such as shape. For image recognition, CNN showed excellent performance. Pre-trained CNN excites informative features of image, it can use for semi-unsupervised feature extraction. The responses of the individual layers to unlabeled input images are learned in every layer. The lower layers of the CNN learned common features such as color, edges. On the other hand, higher-level features are seen in fully connected layers and it reveals their identities.

This paper uses a simple CNN structure that consists of two convolution layers and two fully-connected layer as shown in Fig. 1. Simple CNN is modified from LeNet-5 [21]. The kernel sizes of convolution layers are 32x32, 16x16 and the active function is ReLU. After the iterative learning ends, the trained model is saved. In trained model, activation map extraction block only including convolution layers is uses for semi-supervised feature extraction. The activation vectors of unlabeled images are extracted from the last convolutional layer of pre-trained model.

MD classifies multivariable data points by a distance measure calculated from the data distribution. MD value is calculated by using the normalized value of performance parameters and their correlation coefficients which are the reason for its sensitivity. Therefore, MD is an effective distance metric that measures the distance between a vector and a distribution. In PHM, MD facilitates anomaly detection by determining the system's health. In this paper, the activation values extracted from each test image show a tendency to gradually move away from the distribution of normal conditions as the severity increases. To visualize this trend, this paper defines the health indicator with MD for PMSM condition assessment. The health indicator with MD can be formulated as

$$MD(i) = (a(i)-\mu)\sum\nolimits^{-1}(a(i)-\mu)^T \quad (2)$$

where $MD(i)$ is the MD at the $i^{th}$ data points; $a(i)$ is an absolute activation map of the $i^{th}$ data points; $\mu$ is an average value of activation map; and $\Sigma^{-1}$ is an inverse covariance activation map value.

## IV. CASE STUDY

### A. Experimental Setup

PMSM circuit diagram when inter-turn short circuit fault occurs in the stator winding can be present as Fig. 2. The ratio of short-circuit is defined as the number of turns on one phase divided by the number of short-circuit fault turns. If the short circuit resistance is zero and the short circuit ratio is one, it becomes a complete inter-turn short circuit fault. Therefore, the fault ratio (FR) can be formulated as

$$\text{Fault Ratio (FR)} = \frac{R_f}{R_s} \quad (3)$$

where $R_f$ is by-passing resistance values, and $R_s$ is inter-turn resistance value. This paper uses the motor testbed as shown in Fig. 3. Three current sensor Hioki CT6700 were mounted on the motor power cable. 1.5 kW and 3.0 kW PMSM were used for testing. Three severity of inter-turn short circuit fault are seeded based on FR. The current signal was measured from testbed at a rated rotating speed 3000 rpm for 120 seconds at a sampling rate of 100 kHz. Current signals were collected in an real-field environment with external noise. Therefore, although the collected current signals have no abnormality in driving the motor, there are many spiked components that can cause erroneous diagnosis in the fault diagnosis.

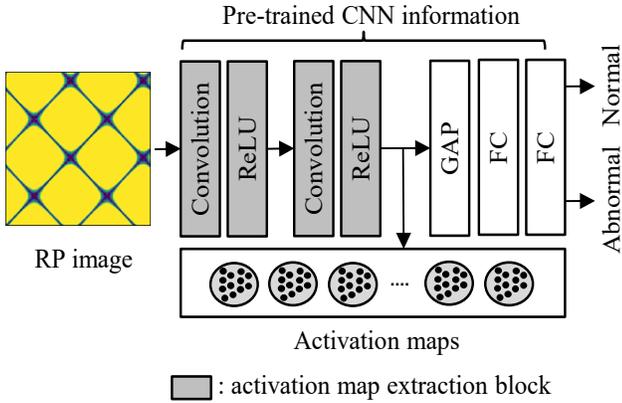

Fig. 1 Unsupervised feature extraction by trained CNN

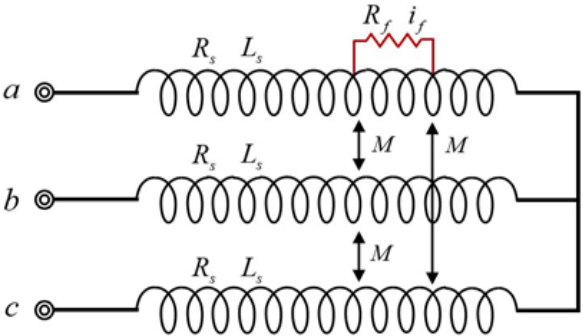

Fig. 2 Stator winding of inter-turn short circuit fault scheme

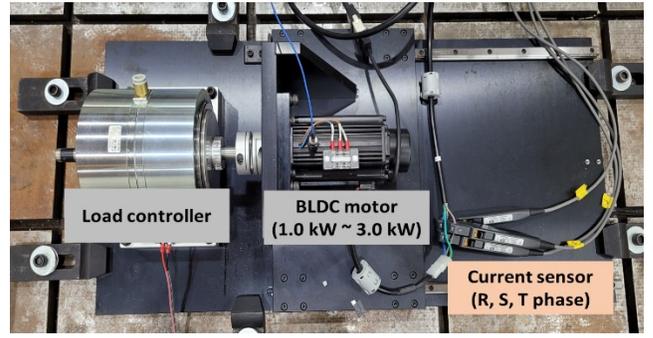

Fig. 3 Description of electric motor testbed

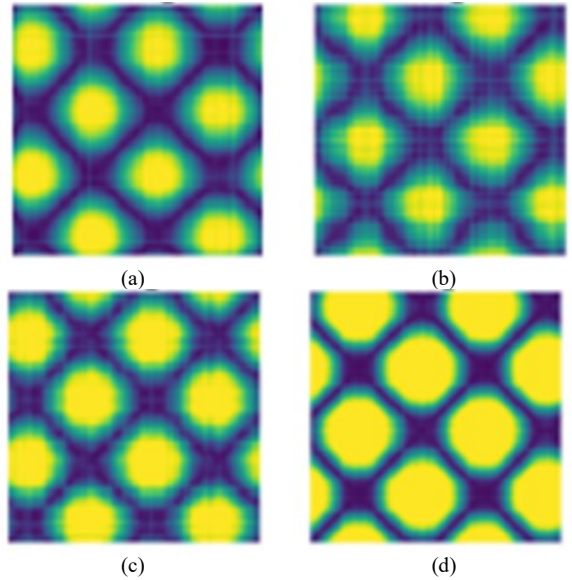

Fig. 4. RP images of stator current signals from 1.5 kW motor: (a) healthy state, (b) faulty state (1.82 %), (c) faulty state (4.33 %), and (d) faulty state (7.80 %)

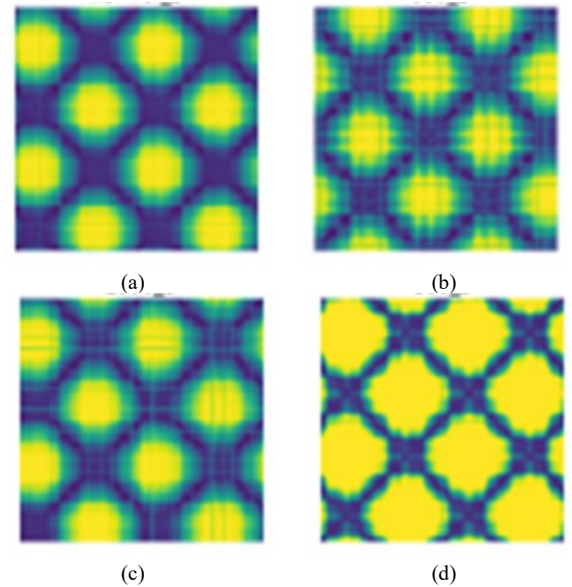

Fig. 5. RP images of stator current signals from 3.0 kW motor: (a) healthy state, (b) faulty state (1.82 %), (c) faulty state (4.33 %), and (d) faulty state (7.80 %)

## B. Results and Discussion

PMSM is hard to classify using only time feature or frequency feature. However, the proposed current signal imaging method shows different patterns on every condition of PMSM as shown in Fig. 4 and Fig. 5. The sinusoidal periodic features of the normal current signal are reflected as a cross pattern in the normal image. Furthermore, the difference in the faulty signal is emphasized in the image as the defect severity increases. The convex pattern has increased as the severity increases. Therefore, proposed method can represent multiple features of motor current signal as a single image. Besides, the patterns of images are same according to the severity regardless of the motor capacity. Even if the images are resized, the features of images do not contaminate well. As a result, the images of dynamic resistance are resized to 95% decrease for fast computation. The resized images are 600 pixels by 600 pixels.

To compare the performance of the proposed health indicator, knowledge-based features are used [17]. The knowledge-based features including maximum value, RMS value, skewness value, kurtosis value, crest factor, and entropy factor are used. The features are normalized to avoid the numerical instability issue. The conventional method is very hard to decide threshold value and the degree of status is not obvious as shown in Fig 6(a). On the contrary, the proposed method clearly shows the status of the motor without feature engineering as shown in Fig 6(b). Therefore, the proposed method can diagnose the motor regardless of the motor specification.

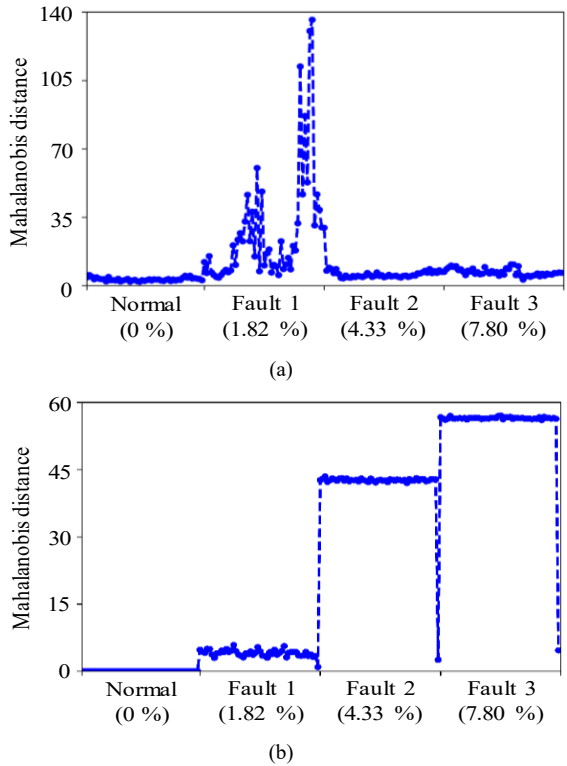

Fig. 6. Health indicator graph: (a) conventional method, and (b) proposed method

## V. Conclusions

This paper presents a PMSM fault diagnosis method using motor current signal. The proposed method consists of three main steps: (1) time-series current signal imaging method using the proposed RP in time-domain, (2) extracting features using convolutional neural networks, and (3) calculating a health indicator using Mahalanobis distance. First, this paper proposes a time-series data imaging method that can detect weak fault features. The proposed time-series data imaging method not only includes all relevant dynamic information of time-series data but also works robustly against data noise and the capacities of the motor. Second, the unlabeled image is transformed into an active value distribution by a pre-trained CNN. Finally, the distance from the normal feature distribution is measured using the Mahalanobis distance. Distance is used to identify the status of PMSM. The proposed method shows the possibility of real-time monitoring of the status of PMSM using current signal. If the image pattern can be distinguished, the image size can be resized for fast processing. Future research can extend the proposed method to identify various types of motor faults and be utilized in other scale motor fault diagnosis studies.

## Acknowledgment

This paper was supported by Korea Institute for Advancement of Technology(KIAT) grant funded by the Korea Government(MOTIE) (P0017006, HRD Program for Industrial Innovation) and also supported by the KAIST AI Institute ("Kim Jae-chul AI Development Fund" AI Dataset Challenge Project) (Project No. N11210251).